\documentclass{iopart}
\usepackage{amsmath,amssymb,bbm,graphicx,color}
\newcommand{\bmsigma}{\boldsymbol \sigma} 
\newcommand{\bmLambda}{\boldsymbol \Lambda}

\def\S{{\sf S}}
\newcommand{\bmlambda}{\boldsymbol \lambda} \def\X{\boldsymbol{X}}
\def\Tr{\hbox{Tr}} 
\begin{document}
\title{Non-Gaussian states by conditional measurements}
\author{Marco G. Genoni$^{1,2}$, 
Federica A. Beduini$^{1}$, Alessia Allevi$^{2}$, Maria Bondani$^{3,4}$, 
Stefano Olivares$^{2,1}$,
Matteo G. A. Paris\footnote{\tt matteo.paris@fisica.unimi.it}$^{1,2,5}$}
\address{$^1$Dipartimento di Fisica, 
Universit\`a degli Studi di Milano, I-20133 Milano, Italy \\
$^2$CNISM, U.d.R. Milano Universit\`a, I-20133 Milano, Italy \\
$^3$National Laboratory for Ultrafast and Ultraintense 
Optical Science, CNR-INFM, I-22100 Como, Italy\\
$^4$CNISM, U.d.R. Como, I-22100 Como, Italy\\
$^5$ISI Foundation, I-10133 Torino, Italy }
\begin{abstract}
We address realistic schemes for the generation of non-Gaussian
states of light based on conditional intensity measurements performed 
on correlated bipartite states. We consider both quantum and classically 
correlated states and different kind of detection, comparing the resulting
non Gaussianity parameters upon varying the input energy and the detection
efficiency. We find that quantum correlations generally lead to
higher non Gaussianity, at least in the low energy regime.
An experimental implementation feasible with current technology is also
suggested.
\end{abstract}
\section{Introduction}
Continuous variable (CV) quantum information has been developed with
Gaussian states and operations, which allow the realization fundamental
protocols as teleportation, cloning and dense coding, and still play a
relevant role in CV quantum information processing.  Recently, also the
non Gaussian (nG) sector of the Hilbert space is receiving attention for
the potential application in entanglement distillation and entanglement
swapping for long distance quantum communication.  In turn, using nG
states and operations, teleportation \cite{Tom,IPS2a,IPS2b} and cloning
\cite{nGclon} of quantum states may be improved. In order to quantify
the amount of nG of a state measures have been recently proposed, based
on \emph{distances} between the quantum state under investigation and a
reference Gaussian state \cite{nGHS,nGRE,nGSimon}.
\par
Gaussian states are generated using linear and bilinear interaction in
optical materials, whereas the generation of nG states requires higher
nonlinearities as those exhibited by a Kerr medium. An alternative
approach is to exploit the effective nonlinearity induced by conditional
measurements. In fact, if a measurement is performed on a portion of a
composite system, the other component is conditionally prepared
according to the outcome of the measurement and the resulting dynamics
may be highly nonlinear. The the rate of success in getting a certain
state is equal to the probability of obtaining a certain outcome and may
be higher than nonlinear efficiency, thus making conditional schemes
possibly convenient even when a corresponding Hamiltonian process
exists.
\par
In this paper we focus on preparation schemes feasible with current
technology and address generation of non-Gaussian states of light by
conditional intensity measurements on the twin-beam state (TWB) generated by
parametric down-conversion and on classically correlated states generated
by mixing a thermal beam with the vacuum in a beam splitter (BS). In particular
we consider different kinds of conditional measurements: ideal photodetection,
inconclusive photodetection and photodetection performed with a finite
quantum efficiency $\eta$ and no dark counts. 
We compare the resulting nG varying all the involved parameters 
and in particular the number of photons
of the initial state.
\par
Let us consider a CV system made of $d$ bosonic modes
described by the mode operators $a_k$, $k=1\dots d$, with
commutation relations $[a_k,a_j^{\dag}]=\delta_{kj}$. A quantum
state $\varrho$ of $d$ bosonic modes is fully described by its
characteristic function $\chi[\varrho](\bmlambda) = \Tr[\varrho\:D(\bmlambda)]$
where $D(\bmlambda) = \bigotimes_{k=1}^d D_k(\lambda_k)$
is the $d$-mode displacement operator, with $\bmlambda =
(\lambda_1,\dots,\lambda_d)^T$, $\lambda_k \in \mathbbm{C}$, and where
$D_k(\lambda_k) =\exp\{\lambda_k a_k^{\dag} - \lambda_k^* a_k \}$ is the
single-mode displacement operator.
The canonical operators are given by
$q_k = (a_k + a^{\dag}_k)/\sqrt{2}$ and
$p_k = (a_k - a_k^{\dag})/\sqrt{2}i$
with commutation relations given by $[q_j,p_k]=i\delta_{jk}$.
Upon introducing the vector $\boldsymbol{R}=(q_1,p_1,\dots,q_d,p_d)^T$,
we have the vector of mean values 
$\X \equiv \X[\varrho]$, ${X}_j = \langle R_j \rangle$, and 
and the covariance 
matrix (CM) $\bmsigma\equiv\bmsigma[\varrho]$ of a quantum state, 
$\sigma_{kj} =\frac{1}{2}\langle R_k R_j + R_j R_k\rangle - \langle R_j
\rangle\langle R_k\rangle$
where
$\langle O \rangle = \Tr[\varrho\:O]$.
A quantum state
$\varrho_G$
is said to be Gaussian if
its characteristic function is Gaussian, that is
$\chi[\varrho_G](\bmLambda) = \exp \left\{- \frac{1}{2}
\bmLambda^T \bmsigma \bmLambda + \X^T\bmLambda \right\}
$,
where $\bmLambda = (\hbox{Re}
\lambda_1, \hbox{Im}\lambda_1, \dots, \hbox{Re} \lambda_d,
\hbox{Im}\lambda_d)^T$. A Gaussian state is fully
determined by CM and $\X$.
\par
For a generic CV quantum state $\varrho$, a measure
of nG based on the quantum relative entropy has been
introduced in \cite{nGRE} as
$\delta[\varrho] = S(\tau) - S(\varrho)$
where $S(\varrho) = - \hbox{Tr}[\varrho \log \varrho]$ is the
von Neumann entropy of a quantum state $\varrho$, and
$\tau$ is the Gaussian state with the same CM and
$\X$ as $\varrho$. 
In this paper we will deal with single-mode non-Gaussian states
that can be written as diagonal mixtures of Fock states
$\varrho = \sum_{n=0}^{\infty} p_n |n\rangle\langle n|$, for
which the reference Gaussian state is the single-mode
thermal state $\nu_N= 1/(1+ N_{ph}) \sum_{n=0}^{\infty}
\left(N_{ph}/(1+N_{ph})\right)^{n} |n\rangle\langle n|$
where $N_{ph} = \hbox{Tr}[\varrho \: a^{\dag}a ]$ is the mean photon
number.
The von Neumann entropy for these states can be easily obtained
and therefore the nG can be written as
\begin{align}
\delta[\varrho] = N_{ph} \log\left(\frac{N_{ph} +1}{N_{ph}}\right)
+ \log ( 1 + N_{ph}) + \sum_{n=0}^{\infty} p_n \log p_n \label{eq:nG}
\end{align}
\section{Non-Gaussian states by conditional measurements}\label{s3}
We want to study the quantum states generated by conditional
measurements on quantum and classically correlated states.
In particular as initial states we consider (i) an entangled
TWB, obtained from (spontaneous)
parametric down-conversion (SPDC) in second-order nonlinear
crystals, (ii) a thermal state mixed with the vacuum state
in a BS of transmissivity $T$. 
The TWB state can be written as
$|\Lambda \rangle\rangle = \sqrt{1 - |\lambda|^2} \sum_{n}
\lambda^n |n\rangle_1 \otimes |n\rangle_2 $ where 
$|n\rangle_j$ denotes the Fock number state
in the Hilbert space of the $j$th mode. The parameter 
$|\lambda|<1$ may be taken as real without loss of
generality and the mean number of photons for each mode 
is given by
$N = \Tr [|\Lambda\rangle\rangle\langle\langle\Lambda|
\: \hat{n}_1 \otimes \mathbbm{1} ]
= \Tr [|\Lambda\rangle\rangle\langle\langle\Lambda|
\: \mathbbm{1} \otimes \hat{n}_2 ]
=  \lambda^2/(1-\lambda^2)$
The output state from a BS of transmissivity $T$
mixing a thermal state $\nu_{2N}$ and a vacuum, is described by a
density operator
\begin{align}
R = \sum_{stpq} T^{(s+t)/2} (1-T)^{(p+q)/2} \nu_{p+s,t+q}
\sqrt{\binom{p+s}{s}\binom{q+t}{q}} |s\rangle\langle t|
\otimes |p\rangle\langle q|
\end{align}
where $\nu_{h,k} = \delta_{h,k} (1+2N)^{-2} [ 2N/(1+2N)]^k$ and $2N$ is 
the mean photon number in $R$. 
In the next subsection we will derive the states generated
from $|X\rangle\rangle$ and $R$ by means of different
conditional measurements, that is
\begin{align}
\varrho_{\Lambda,i} &= \frac{\Tr_2[|\Lambda\rangle\rangle\langle\langle
\Lambda| \: \mathbbm{1}\otimes M^{(i)}]}{\Tr[|\Lambda\rangle\rangle\langle\langle
\Lambda| \: \mathbbm{1}\otimes M^{(i)}]}\qquad
\varrho_{R,i}=\frac{\Tr_2[R \: \mathbbm{1}\otimes M^{(i)}]}{\Tr[R \: \mathbbm{1}\otimes M^{(i)}]} \nonumber
\end{align}
where $i$ denotes the kind of measurement we are performing and $M^{(i)}$ is the operator of the corresponding
probability operator valued measure (POVM). In particular we will
consider $(1)$ ideal photodetection, $(2)$ (ideal)
inconclusive photodetection and $(3)$ photodetection with
a non-unit quantum efficiency. We will then evaluate their
non-G parameter and describe their photon statistics.
In particular we will give the mean photon number $N_{ph}$,
the variance $\sigma_{ph}^2$ and the corresponding Fano Factor
defined as $F_{ph} = \sigma_{ph}^2 / N_{ph}$.
\subsection{Ideal photodetection}
The POVM
of an ideal photon-resolving detector is given by projectors
on the Fock number basis: $P_m = |m\rangle\langle m|$.
By taking the TWB $|\Lambda\rangle\rangle$ as the initial state,
the output state after measuring $m$ photons is simply the
corresponding Fock state $\varrho_{\Lambda,1}=
|m\rangle\langle m|$.
The photon statistics of
$\varrho_{\Lambda,1}$ is
trivial, since $N_{ph}=m$ and
both $\sigma_{ph}^2$ and
$F_{ph}$ are null. The nG of the output state can be easily
evaluated by using Eq.~(\ref{eq:nG}) and
it  monotonically depends on the number of
detected photons $m$.\par
\begin{figure}[ht!]
\centerline{\includegraphics[width=0.35\textwidth]{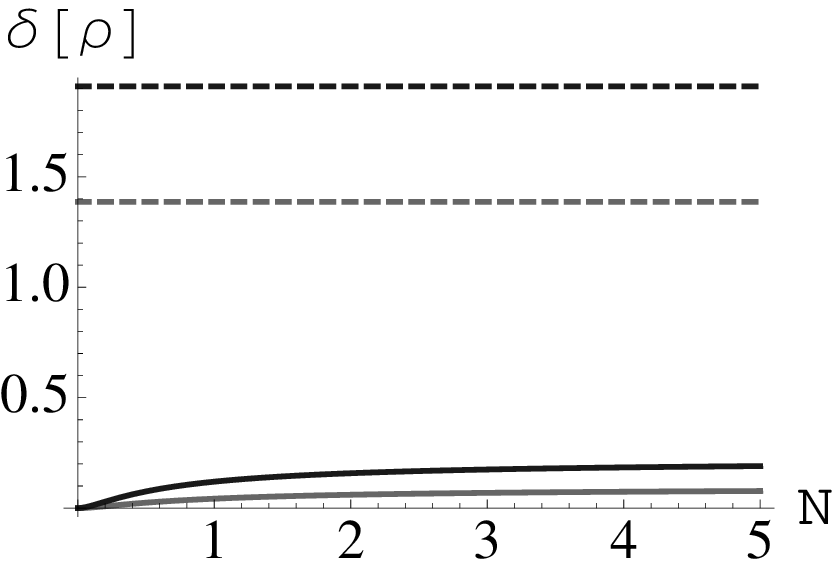}
\includegraphics[width=0.35\textwidth]{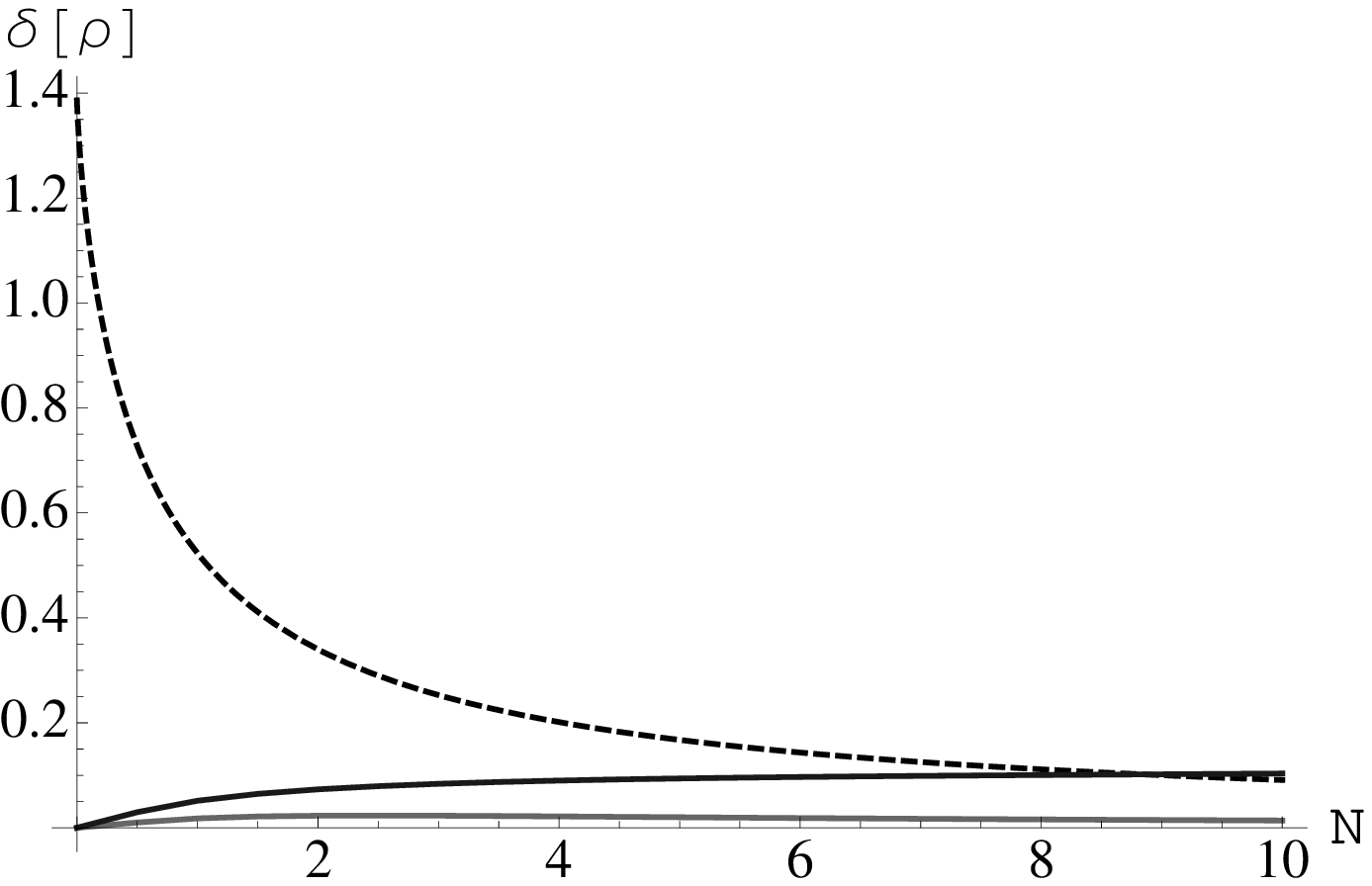}}
\caption{Left: Non-Gaussianity of quantum states generated by performing
an ideal photodetection on one arm of a TWB $|\Lambda\rangle\rangle$
(dashed lines) or of the classically
correlated state $R$ (solid lines) obtained with a
BS with transmissivity $T=0.9$,
as a function of the initial average
number of photons $N$ and for different values of detected
photons $m$. From lighter to darker gray:
 $m=\{1,2 \}$.
Right: Non-Gaussianity of quantum states generated by inconclusive
photo-detection on a TWB $|\Lambda\rangle\rangle$
 (black dashed line) or on the quantum
state $R$ (solid lines) for different values of the
BS transmissivity $T$ 
(from lighter to dark gray: $T=\{0.5, 0.99 \}$)
as a function of the initial number of photons $N$.
\label{f1}
}
\end{figure}

In the case of the classically-correlated state $R$ with
$N$ mean photons in each mode, performing an ideal photodetection 
on one mode leaves the other mode in the following conditional state:
\begin{align}
\varrho_{R,1} &= \left( \frac{1+2N(1-T)}{1+2N} \right)^{m+1}
\sum_{s=0}^{\infty} \binom{m+s}{m} \left(
\frac{2 N T}{1+2N} \right)^s
|s\rangle\langle s| \nonumber
\end{align}
In this case the mean
number of photons, the variance and the Fano factor of the ouput state
are non-trivial but can be evaluated obtaining the following results
$$
N_{ph} = \frac{2(1+m)\:N\:T}{1+2 N (1-T)} \qquad
\sigma_{ph}^2 = \frac{ 2 (1+m) N T (1+2N)}{(1+2N(1-T))^2}  
$$
$$
F_{ph} = \frac{1+2N}{1+2N(1-T)}. 
$$
The corresponding nG $\delta[\varrho_{R,1}]$ can be evaluated numerically
and is plotted in Fig.~\ref{f1} along with the nG
of the previous case.
As one can observe from the plot,
the state generated by the classically-correlated state is less
non-Gaussian than that generated by TWB.
Its nG increases with the number of detected photons $m$, with
the initial mean number of photons $N$ and with the transmissivity
of the BS $T$. Though increasing with $N$, we
have numerical evidences that the asymptotic value
for $N\rightarrow \infty$,
is below the nG of the Fock states
$|m\rangle\langle m|$ obtained from the TWB.
\subsection{Inconclusive photodetection}
The ideal inconclusive photodetection is described by
the POVM operator $\Pi_{off}=|0\rangle\langle 0|$, if no photons
are detected, and $\Pi_{on}=\mathbbm{1}-|0\rangle\langle 0|$
if one or more photons are detected. Let us first consider
the state obtained when one ore more photons are
detected on one mode of the TWB $|\Lambda\rangle
\rangle$. The state that is generated can be written in the
Fock basis as
\begin{align}
\varrho_{\Lambda,2} &= \frac{1 - \lambda^2}{\lambda^2} \sum_{s=1}^{\infty}
\lambda^{2s} |s\rangle\langle s|. \nonumber
\end{align}
Note that, for $\lambda\rightarrow 0$ ($N\rightarrow 0$)
we obtain the Fock state $|1\rangle\langle 1|$, while
for $\lambda\rightarrow 1$ ($N\rightarrow \infty$),
the state is not physical.
The photon statistics of $\varrho_{R,2}$ can be summarized
in terms of the initial number of photons $N$ using the parameters
$N_{ph} = 1+N$, $\sigma_{ph}^2 = N (1+N)$, and $F_{ph} = N$. 
The corresponding nG $\delta[\varrho_{\Lambda,2}]$
can be evaluated numerically and is plotted in Fig.~\ref{f1}.
It is apparent from the plot that $\delta[\varrho_{\Lambda,2}]$
is a monotonically decreasing function of the initial number of
photons $N$, starting from the nG of the
Fock state $|1\rangle\langle 1|$ and approaching zero
for $N\rightarrow\infty$.\par
Let us consider now the conditional state generated from the
classically-correlated state $R$. It is diagonal in the
Fock basis and can be written as
\begin{align}
\varrho_{R,2}
&= \frac{1+2N(1-T)}{2N(1-T)} \sum_{s=0}^{\infty}
\left( \frac{2NT}{1+2N} \right)^s
\left[ \frac{1}{1+2NT} \left( \frac{1+2N}{1+2NT} \right)^s -
\frac{1}{1+2N} \right] |s\rangle\langle s| \nonumber
\end{align}
The average photon number, its variance and the Fano factor are
$$
N_{ph} = \frac{4N\:T (1+N(1-T))}{1+2N(1-T)} \qquad
\sigma_{ph}^2 = \frac{4\:N\:T (1 + N (3-T+4 T\: N^2 (1-T)^2 +2 N (1-T^2)))}
{(1+2N(1-T))^2} 
$$
$$
F_{ph} = 2(1+N\:T) - \frac{2(1+N)}{1+N(1-T)} + \frac{1+2N}{1+2N(1-T)}
$$
The nG $\delta[\varrho_{R,2}]$ can be numerically evaluated
and is plotted in Fig.~\ref{f1}. In this case it has not a
monotonic behavior as a function of $N$ while it is still
an increasing function of the transmissivity $T$.
In the low energy regime the state generated starting with
a TWB is definitely more nG
than that generated by means of $R$. By increasing
the initial mean number of photons, since $\delta[\varrho_{R,1}]$
approaches zero, we observe a region where the classical
correlated state gives birth to a more non-Gaussian state
than the quantum one.
\subsection{Inefficient photodetection}
\begin{figure}[ht!]
\centerline{\includegraphics[width=0.35\textwidth]{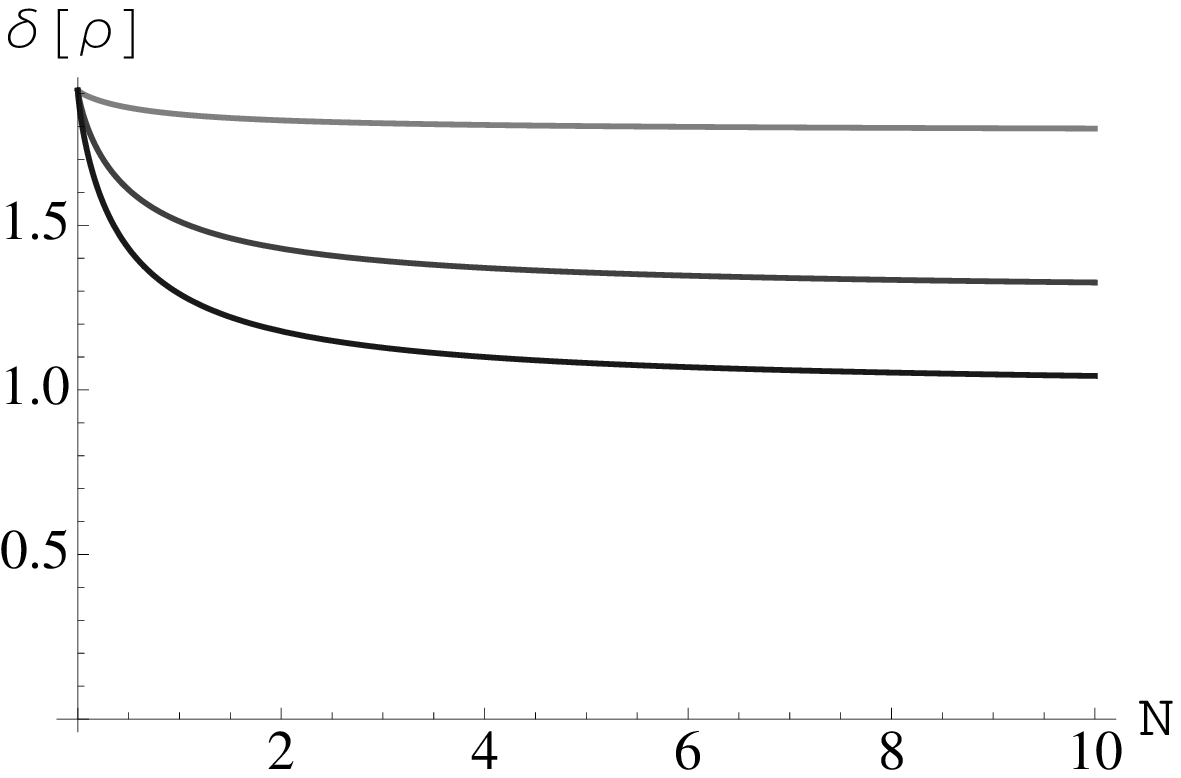}
\includegraphics[width=0.35\textwidth]{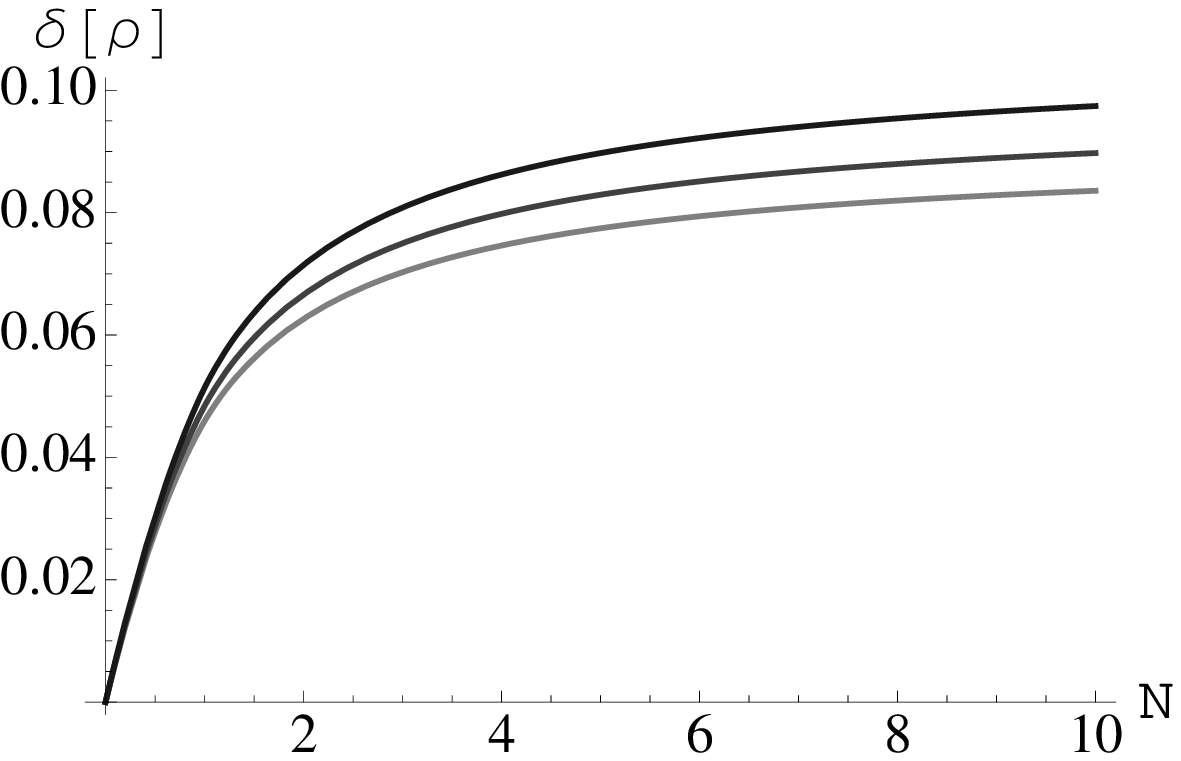}}
\caption{nG of conditional states generated by performing
an inefficient photodetection on one mode of a TWB
$|\Lambda\rangle\rangle$ (left figure) or of the classical
correlated state $R$ (right figure) obtained by a BS
with transmissivity $T=0.5$,
as a function of the initial average
number of photons $N$, for $m=2$ photons detected
and for different values of the efficiency $\eta$.
From darker to ligher gray: $\eta=\{ 0.8, 0.9, 0.99 \}$.
As one can observe comparing the two plots, the nG 
of states generated from TWB are higher 
than the one obtained from thermal states.
\label{f2:ineff}
}
\end{figure}

The POVM of a photon-number resolving detector with
quantum efficiency $\eta$ and no dark counts is given by
the Bernoullian convolution of the ideal number
projectors $P_l=|l\rangle\langle l|$, and thus,
for $m$ detected photons, it is described by the operator 
$\Pi_m = \eta^m \sum_{l=m}^{\infty} (1-\eta)^{l-m}
\binom{l}{m} P_l 
$.
The state generated by detecting $m$ photons on
one arm of a TWB can be thus written as
\begin{align}
\varrho_{\Lambda,3}
&= \frac{1-\lambda^2(1-\eta)}{(\lambda^2 (1-\eta))^m}
\sum_{l=m}^{\infty} \binom{m}{l} (\lambda^2(1-\eta))^l |l\rangle
\langle l |. \nonumber 
\end{align}
Here we summarize the photon statistics of this state
in terms of the previously introduced quantities, by substituting
$\lambda$ with $N$ 
$$
N_{ph} = \frac{(1+m)(1+N)}{1+N\eta} - 1  \qquad
\sigma_{ph}^2 = \frac{(1+m)(1-\eta) N (1+N)}{(1+\eta N)^2}
$$
$$
F_{ph} = \frac{N(1+N)(1+m)(1-\eta)}
{(m(1+N) + N(1-\eta))(1+N\eta)} 
$$
The nG can be evaluated numerically and it is
plotted in Fig.~\ref{f2:ineff} for different values of the
parameter. We observe that, as expected, nG is
a monotonically increasing function of the number of detected photons
$m$. Moreover
we observe that it decreases with the initial mean photon
number $N$, while it increases at increasing detector efficiency. By choosing $\eta=1$ we trivially obtain
the results of the ideal photodetection.
For $N\rightarrow\infty$, the state generated by measuring
a TWB turns out to be the corresponding (normalized) POVM operator,
that is $\varrho_{\Lambda,3}= \eta \Pi_m$. The asymptotic values approached
for $N\rightarrow\infty$ correspond to
the nG of this particular state, and increases
at increasing $\eta$. 
Finally we consider the classically-correlated state $R$. Also
in this case we can obtain the state generated
by measuring $m$ photons in one arm, with efficiency $\eta$
\begin{align}
\varrho_{R,3}
&= \left(\frac{1+2N\eta(1-T)}{1+2N(T+\eta(1-T))}\right)^{m+1}
\sum_{s=0}^{\infty} \binom{m+s}{m}
\left(\frac{2 N T}{1+2N(T+\eta(1-T))}\right)^{s} |s\rangle\langle s| \nonumber
\end{align}
The mean number of photons, its variance and Fano
factor result
$$
N_{ph} = \frac{2 (1+m) N T}{1+2N\eta (1-T)}  \qquad
\sigma_{ph}^2 = \frac{2 (1+m) N T (1+2N(T+\eta(1-T)))}
{(1+2N\eta(1-T))^2}
$$
$$
F_{ph} = 1 + \frac{2NT}{1+2N\eta(1-T)}. 
$$
Again the nG $\delta[\varrho_{R,3}]$
has been numerically evaluated and is plotted
in Fig.~\ref{f2:ineff}. As in the ideal case, the
nG obtained from a classically-correlated
state take lower values than those from TWB.
Moreover we again observe that a higher nG
is obtained at higher transmissivity (as in the
previous cases) and, unlike the TWB case,
at lower detection efficiency and higher
values of the initial number of photons $N$.
For $N\rightarrow\infty$ the asymptotic state
can be written as
\begin{align}
\varrho_{R,3} = \left( \frac{\eta (1-T)}{T+\eta (1-T)}
\right)^{m+1} \sum_{s=0}^{\infty} \binom{m+s}{m} \left(
\frac{T}{T+\eta (1-T)} \right)^s |s\rangle\langle s|.
\end{align}
Again we have numerical evidences that for $N\rightarrow\infty$
the nG obtained for the classical correlated state
are below those obtained by measuring the TWB state.
\section{Experimental proposal}\label{s4}
The experimental implementation of the optical states described in the
previous sections, namely TWB states and thermal states, would allow us
to verify the correctness of our model. Obviously, as the real detectors
are endowed with a non-ideal quantum efficiency, here we consider only
the case of imperfect detection presented in Sec.~\ref{s3}.  In order to
match the requirements of the theoretical model we plan to use two
hybrid photodetectors (HPD, Hamamatsu) as the detectors, which are
endowed with a partial photon-number resolving capability and no dark
counts \cite{JMO}.  \par First of all, we discuss the case of the
generation of TWB states with a sizeable mean photon number. Such states
can be obtained by pumping a second-order nonlinear crystal (either in
type-I or type-II phase-matching) by means of a pulsed laser in a
travelling-way configuration \cite{jointdiff}.
\begin{figure}[ht!]
\centerline{\includegraphics[angle=270,width=0.5\textwidth]{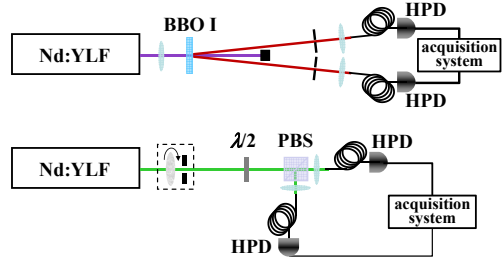}}
\caption{Proposed experimental setups. Upper panel: 
TWB state. Lower panel: classically correlated thermal state.
\label{f:setup}
}
\end{figure}
As sketched in the upper panel of Fig.~\ref{f:setup}, we can exploit
either the third harmonics ($\lambda=349$ nm, 4.4-ps-pulse duration) or
the fourth harmonics (@ $\lambda=262$ nm, 4-ps-pulse duration) of a
frequency-tripled Nd:YLF mode-locked laser amplified at 500 Hz to
produce bright cones of spontaneous parametric down-conversion in a
type-I $\beta$-BaB$_2$O$_4$ crystal. The TWB state will be
measured by selecting with suitable pin-holes a pair of twin coherence
areas on the signal and idler cones \cite{subshot}. In order to make the
quantum efficiencies of the two arms as equal as possible, we will
operate at frequency degeneracy. Note that with both choices for the
pump, the measured TWB wavelength will be in the visible spectrum
range, where most photodetectors have the maximum quantum efficiency.
Multi-mode fibers having a good transmissivity in the visible spectrum
range will be employed to collect all the light passing the pin-holes
and to send it to the detectors. The study of the dependence of the nG
factor on the overall quantum efficiency will be experimentally
reproduced by inserting a variable neutral density filter on the arm of
the detector performing conditional measurement.  Unfortunately,
depending on crystal length and phase matching conditions, the TWB
state produced by ps-pulsed lasers is intrinsically temporal multi-mode.
For this reason, the single-mode theory previously developed should be
extended to correctly describe the experimental situation.  \par The
production of a single-mode pseudo-thermal state can be easily obtained
by inserting a rotating ground glass plate on the pathway of a coherent
field, followed by a pin-hole selecting a single coherence area in the
far-field speckle pattern \cite{ASL} (see the lower panel of
Fig.~\ref{f:setup}). To match the maximum quantum efficiency of the
detectors, we can use the second harmonics ($\lambda= 523$ nm,
5.4-ps-pulse duration) of the Nd:YLF laser described above. The thermal
light will then be split in two parts by means of a BS in
order to perform conditional measurements. To make the setup more
versatile and in particular to have the possibility to study the
dependence of the nG factor on the transmissivity of the
BS, it is convenient to substitute the ordinary BS
with a system composed by a half-wave plate ($\lambda$/2 in
Fig.~\ref{f:setup}) and a polarizing cube BS. Again the
measurements can be performed with a pair of hybrid photodetectors.
\section{Conclusions}\label{s5}
We have suggested an experimentally feasible scheme for the generation 
of non-Gaussian states of light by conditional intensity measurements 
on correlated bipartite states. We have considered both quantum and classically 
correlated states and evaluated the amount of resulting nG for 
different kind of detection and input energy. Although the emergence
of nG is not related to the quantum nature of the involved beams
we found that quantum correlations generally lead to
higher nG, especially in the low energy regime.

\end{document}